\RequirePackage{fix-cm}

\documentclass[twocolumn]{svjour3}
\smartqed
\usepackage{graphicx}
\usepackage{url}
\usepackage{marvosym}
\usepackage{amssymb}
\usepackage{color}

\journalname{Brazilian Journal of Physics}
\begin{document}

\title{The Gravitational Wave Background From Coalescing Compact Binaries: A New Method}

\titlerunning{ }

\author{Edgard F. D. Evangelista \and Jos\'{e} C. N. de Araujo}

\authorrunning{ }

\institute{E. F. D. Evangelista \and J. C. N. de Araujo (\Letter)\at
              Instituto Nacional de Pesquisas Espaciais -- Divis\~{a}o de Astrof\'{i}sica, Av. dos Astronautas 1758, S\~{a}o Jos\'{e} dos Campos, 12227-010 SP, Brazil \\
%              Tel.: +55-12-32087214\\
              \email{jcarlos.dearaujo@inpe.br}
           \and
            E. F. D. Evangelista
            \at
              \email{edgard.evangelista@inpe.br}
}

\date{Received: date / Accepted: date}

\maketitle

\begin{abstract}

Gravitational waves are perturbations in the spacetime that propagate at the speed of light. The study of such phenomenon is interesting because many cosmological processes and astrophysical objects, such as binary systems, are potential sources of gravitational radiation and can have their emissions detected in the near future by the next generation of interferometric detectors. Concerning the astrophysical objects, an interesting case is when there are several sources emitting in such a way that there is a superposition of signals, resulting in a smooth spectrum which spans a wide range of frequencies, the so-called stochastic background. In this paper, we are concerned with the stochastic backgrounds generated by compact binaries (i.e. binary systems formed by neutron stars and black holes) in the coalescing phase. In particular, we obtain such backgrounds by employing a new method developed in our previous studies.

%\PACS{04.30.Db \and 95.85.Sz}
\end{abstract}

\section{Introduction}
\label{intro}

Binary systems are among the best known sources of gravitational waves. Specially interesting are the double neutron star systems (NSNS systems), the binaries formed by a neutron star and a black hole (BHNS systems) and the binaries of black holes (BHBH systems), because their emissions have a high probability of being detected in the near future.

Besides, it is well known that binaries lose energy and momentum via emission of gravitational waves, which cause reductions in their orbital distances, consequently increasing the orbital frequencies. Moreover, for systems in circular orbits, the frequency of the emitted waves is twice the orbital frequency. The process continues until the systems reach the coalescing phase, where the systems leave the periodic regime and start the merging phase.

On the other hand, concerning the study of the gravitational radiation itself, the stochastic backgrounds are of special interest. Backgrounds can be generated when, for example, there is a superposition of signals of several sources, resulting in smooth-shaped spectra spanning a wide range of frequencies. In particular, we are concerned in this paper with the backgrounds generated by a population of coalescing compact binaries formed from redshifts ranging from $z=0$ up to $z\sim 20$, i.e., cosmological binaries.

We can find in the literature some very interesting works on this issue, such as \cite{regim,regim2,regim3,wu}, where the authors calculated the backgrounds generated by coalescing NSNS systems. Generally speaking, they used Monte Carlo techniques to simulate the extragalactic population of compact binaries. Besides, these authors considered the time evolution of the orbital frequency by using the ``delay time" (that is, the interval of time between the formation and the coalescence of the systems) in their calculations. In our method, we consider the time evolution in an explicit form, where the equation describing the evolution of the orbital frequency is taken at each instant of life of a given binary.

In Zhu et al.\cite{zhu}, the authors consider the spectra generated by coalescing BHBH systems. In this paper, they assume average quantities for the energy emissions of single sources; on the other hand, in our calculations we consider all the values for the orbital parameters a system can have, in the form of distribution functions.

Still concerning the spectra generated by coalescing BHBH systems, Marassi et al. \cite{marassi} adopted an updated version of the SeBa\footnote{see \url{www.sns.ias.edu/~starlab}} population synthesis code, in which the masses of the black holes range from $\sim 6\mbox{M}_{\odot}$ and $\sim 20\mbox{M}_{\odot}$.

Roughly speaking, the various papers cited above show results that are characterized by spectra with frequencies ranging from $\sim 10$Hz and $\sim 10^{3}$Hz and with maximum amplitudes located in the interval ranging from $\sim 400$Hz to $\sim 800$Hz. As shown below, the backgrounds we generated have similar forms to the ones we mentioned, though our results show, in general, higher amplitudes. This difference will be discussed timely.

Further, as it will be shown, our method  has the very useful characteristic of being numerically simple, in the sense that it does not demand a heavy computational work.

In this paper, the spectra generated by the coalescing binaries will be calculated by means of\cite{araujo05,araujo00}

\begin{equation}
h_{\mbox{\scriptsize{BG}}}^{2}=\frac{1}{\nu_{\mbox{\scriptsize{obs}}}}\int h_{\mbox{\scriptsize{source}}}^{2}dR ,
\label{bg}
\end{equation}
where $h_{\mbox{\scriptsize{BG}}}$ represents the dimensionless amplitude of the spectrum, $\nu_{\mbox{\scriptsize{obs}}}$ is the observed frequency, $h_{\mbox{\scriptsize{source}}}$ is the amplitude of the signals generated by each source and $dR$ is the differential rate of generation of gravitational waves. It is worth pointing out that (\ref{bg}) was deduced from an energy-flux relation. In fact, in a paper by de Araujo et al.\cite{araujo00}, the authors gave a detailed derivation of this equation, showing its robustness. Actually, one can use (\ref{bg}) in the calculation of different types of stochastic backgrounds, provided that one knows $h_{\mbox{\scriptsize{source}}}$ and the corresponding $dR$ to the case one is dealing with.

Here, $h_{\mbox{\scriptsize{source}}}$ has the form\cite{evans,thorne}

\begin{eqnarray}\nonumber
h_{\mbox{\scriptsize{source}}}&=&7.6\times 10^{-23} \\ &&\times \left(\frac{\mu}{M_{\odot}}\right)\left(\frac{M}{M_{\odot}}\right)^{2/3}\left(\frac{1\mbox{Mpc}}{d_{\mbox{\scriptsize{L}}}}\right)\left(\frac{\nu}{1\mbox{Hz}}\right)^{2/3} ,
\label{source}
\end{eqnarray}
where $\nu$ is the emitted frequency that, in this case, is the frequency emitted by a coalescing system (note that $\nu$ and $\nu_{\mbox{\scriptsize{obs}}}$ are related to each other by \ref{redshift}), $\mu$ is the reduced mass and $M$ is the total mass. The differential rate $dR$ in writing in following form

\begin{equation}
dR=\frac{dR}{dV}\frac{dV}{dz}f_{1}(m_{1})f_{2}(m_{2})dzdm_{1}dm_{2} ,
\label{dr}
\end{equation}
where $dV$ is the comoving volume element, $z$ is the redshift and $f_{1}(m_{1})$ and $f_{2}(m_{2})$ are the mass distribution functions of the components of the systems. In fact, for neutron stars such distributions are given by Dirac's delta functions, since we are considering that all these objects have the same mass of $1.4\mbox{M}_{\odot}$; for black holes we use the function\cite{feryal}

\begin{equation}
\label{blackhole}
f(m)=0.332\hspace{1pt}\mbox{exp}\left[-0.347(m-7.8)^{2}\right]\mbox{,}
\end{equation}
where $m$ is given in solar mass units.

The term $dV/dz$ is known from Cosmology, and the problem of determining $dR$ comes down to the calculation of $dR/dV$.

At this point, one could ask in what the present study differs from the previous ones. We will see that the main difference has to do with the application of a new method to calculate $dR$ developed in our previous papers\cite{edgard13,edgard14}.

The paper is organized as follows: in Section \ref{sec2}, we show the main steps to obtain $dR$; then, with (\ref{bg}), $dR$ and (\ref{source}) at hand, we calculate $h_{\mbox{\scriptsize{BG}}}$ for the three families of compact binaries, which are explained in Section \ref{sec3}; in Section \ref{sec4} we present the results and discuss, in particular, the detectability of the backgrounds studied here by the interferometric detectors Laser Interferometer Space Antenna (LISA) (now evolved LISA (eLISA)), Big Bang Observer (BBO),
DECI-Hertz Interferometer Gravitational wave Observatory
(DECIGO), Advanced Laser Interferometer Gravitational
wave Observatory (ALIGO), Einstein Telescope (ET), and
the cross-correlation of pairs of ALIGOs and ET; finally, in Section \ref{sec5} we present our conclusions.

\section{Calculation of $dR/dV$}
\label{sec2}

We present here the main steps for the calculation of $dR/dV$ (we refer the reader to Refs. \cite{edgard13,edgard14} for details).

We start by writing $dR/dV$ in the form

\begin{equation}
\label{dr2}
\frac{dR}{dV}=\frac{d\nu}{dt}\int n_{\mbox{\scriptsize{bin}}}H(\nu,t,t_{0})dt_{0} ,
\end{equation}
where the expression in the right-hand side is the formation rate of systems per comoving volume that reach the frequency $\nu$ at the instant $t$ (see the derivation in Appendix \ref{apenA} and also in Ref.\cite{edgard13}). Also, $t_{0}$ refers to the instant of birth of the systems.

In (\ref{dr2}), $H(\nu,t,t_{0})$ is the frequency distribution function of the binaries, which has the form (we refer the reader to Appendix \ref{apenB} and Ref.\cite{edgard13} for the derivation of this equation):

\begin{eqnarray}\nonumber
H(\nu)&=&\frac{2C}{3}\\
&&\times\left[\frac{GM\pi^{2}}{4}\right]^{1/3}\nu^{-11/3}\nu_{0}^{2}\mbox{exp}\left[\frac{-(r-\bar{r})^{2}}{2\sigma^{2}}\right],
\label{dist2}
\end{eqnarray}
where $\nu_{0}$ is the initial frequency; $C$, $\sigma$ and $\bar{r}$ are constants given in Table \ref{tab2}; and $r$ is the orbital distance, related to $\nu$ by means of Kepler's third law.

On the other hand, $n_{\mbox{\scriptsize{bin}}}$ is the binary formation rate density that, for NSNS systems is given bys

\begin{equation}
\label{bfrd}
n_{\mbox{\scriptsize{nsns}}}=\lambda_{\mbox{\scriptsize{nsns}}}\frac{\dot{\rho}_{\ast}(z_{0})}{1+z_{0}}\hspace{5pt}\mbox{yr}^{-1}\mbox{Mpc}^{-3}.
\end{equation}
Here, $z_{0}$ refers to the redshift of birth of the systems, which is related to $t_{0}$
via the usual expression that can be found in any textbook of cosmology; and $\dot{\rho}_{\ast}(z_{0})$ is the star formation rate density (SFRD).

There are, in the literature, many different proposals to the SFRD, although they do not differ from each other very significantly. Here we adopt, as a fiducial one, that given by Springel and Hernquist\cite{springel}, namely

\begin{equation}
\dot{\rho}_{\ast}(z)=\dot{\rho}_{m}\frac{\beta e^{\alpha(z-z_{m})}}{\beta-\alpha+\alpha e^{\beta(z-z_{m})}}\hspace{5pt}\mbox{M}_{\odot}\mbox{yr}^{-1}\mbox{Mpc}^{-3} ,
\label{starfor}
\end{equation}
where $\alpha=3/5$, $\beta=14/15$, $z_{m}=5.4$ and with $\rho_{m}=0.15\mbox{M}_{\odot}\mbox{yr}^{-1}\mbox{Mpc}^{-3}$ fixing the normalization.

Besides, in (\ref{bfrd}), $\lambda_{\mbox{\scriptsize{nsns}}}=\beta_{\mbox{\scriptsize{ns}}}f_{p}\Phi_{\mbox{\scriptsize{ns}}}$ (see \cite{regim}) is the mass fraction of stars that is converted into neutron stars, where $\beta_{\mbox{\scriptsize{ns}}}$ is the fraction of binaries that survive to the second supernova event; $f_{p}$ gives the fraction of massive binaries (that is, those systems where both components can generate supernovae) and $\Phi_{\mbox{\scriptsize{ns}}}$ is the mass fraction of progenitors that originates neutron stars, which, in the present case, is calculated by

\begin{equation}
\Phi_{\mbox{\scriptsize{ns}}}=\int^{25}_{8}\phi(m)dm
\end{equation}
where $\phi(m)=Am^{-(1+x)}$ is the Salpeter mass distribution\cite{salpeter} with $x=1.35$ and $A=0.17$. Numerically, we have $\beta_{\mbox{\scriptsize{ns}}}=0.024$, $f_{p}=0.136$ and $\Phi_{\mbox{\scriptsize{ns}}}=5.97\times 10^{-3}\mbox{M}_{\odot}^{-1}$.

Now, considering (\ref{bfrd}), we can write (\ref{dr2}) in the form

\begin{equation}
\frac{dR}{dV}=\lambda_{\mbox{\scriptsize{nsns}}}\dot{\rho}_{\ast,c}(z) ,
\end{equation}
where, following the notation adopted by Zhu et al\cite{zhu}, $\dot{\rho}_{\ast,c}(z)$ is calculated by

\begin{equation}
\dot{\rho}_{\ast,c}(z)=\int \frac{\dot{\rho}_{\ast}(z_{0})}{1+z_{0}}\left[H(\nu,t,t_{0})\frac{d\nu}{dt}\right]dt_{0}.
\end{equation}

In this context, following Wu et al\cite{wu}, $\lambda_{\mbox{\scriptsize{nsns}}}$ is given by $\lambda_{\mbox{\scriptsize{nsns}}}=r_{o,\mbox{\scriptsize{nsns}}}/\dot{\rho}_{\ast,c}(0)$ where $r_{o,\mbox{\scriptsize{nsns}}}$ is the local coalescence rate and $\dot{\rho}_{\ast,c}(0)$ has the form

\begin{equation}
\dot{\rho}_{\ast,c}(0)=\int\frac{\dot{\rho}_{\ast}(z_{0})}{1+z_{0}}\left[H(\nu,t,t_{0})\frac{d\nu}{dt}\right]_{z=0}dt_{0}.
\end{equation}

For BHNS and BHBH systems, we have similar expressions, but with different values of $n_{\mbox{\scriptsize{bin}}}$. Such values can be estimated by means of the results found in Belczynski et al.\cite{belczynski}, where these authors claim that the population of binaries is formed by $61$\% of NSNS, $30$\% of BHBH and $9$\% of BHNS binaries. Therefore, $n_{\mbox{\scriptsize{bhns}}}$ and $n_{\mbox{\scriptsize{bhbh}}}$ can be related to $n_{\mbox{\scriptsize{nsns}}}$ by means of these proportions.

It is worth mentioning that Belczynski et al.\cite{belczynski} studied compact binaries with merger times lower than $10^{10}$yr; that is, they considered coalescing binaries. Basically, in the simulations they used, the binaries are formed through several different channels. Specifically, NSNS systems are formed through $14$ different channels, where there is a predominance of channels containing hypercritical accretion between a low-mass helium giant and its companion neutron star. On the other hand, BHNS and BHBH systems are formed through just for and three channels, respectively, where there is a moderate predominance of mass transfer events.

\section{Calculation of the spectra}
\label{sec3}

In this Section, we present the calculations of the spectra generated by NSNS, BHBH and BHNS systems. Although we are using (\ref{bg}) for the three cases, the calculations are different for each family of binaries. Therefore, we will show the calculations separately.

\subsection{NSNS systems}

Neutron stars, according to theories of stellar evolution and observations (see, e.g., Ref. \cite{ostlie}) have characteristic masses that fall in a narrow interval around $1.4\mbox{M}_{\odot}$. So, in this paper we are considering that all neutron stars have masses of $1.4\mbox{M}_{\odot}$, which is a realistic choice, and at the same time, a simplification.

One could ask how the results would be affected by the choice of the neutron star (NS) equation of state (EOS). Since we are considering the coalescing phase, the results depend mainly on the mass of the NSs. NSNS systems are characterized by a specific coalescence frequency that, according to, e.g., Ref. \cite{iscoNSNS}, may be considered as $\approx 900$Hz for a pair of two $1.4\mbox{M}_{\odot}$ NSs. The choice of the EOS is certainly important for the subsequent phase of evolution of the system, when the merger phase takes place.

Although there is just a value for the coalescence frequency, the spectrum will be spread over a wide range of frequencies. This behavior is due to the cosmic redshift, because systems emitting at the same frequency, but at different redshifts, will generate signals with different observed frequencies, obeying
\begin{equation}
\label{redshift}
 \nu_{\mbox{\scriptsize{obs}}}=\frac{\nu}{1+z}
\end{equation}
where, in this case, we are considering $\nu=900$Hz. Moreover, as we are considering that the redshift has the minimum and maximum values given by $z_{\mbox{\scriptsize{min}}}=0$ and $z_{\mbox{\scriptsize{max}}}=20$, respectively, the observed frequencies will have minimum and maximum values of $\nu^{\mbox{\scriptsize{min}}}_{\mbox{\scriptsize{obs}}}=42.86$Hz and $\nu^{\mbox{\scriptsize{max}}}_{\mbox{\scriptsize{obs}}}=900$Hz, respectively.

Since the masses of the components of the systems and their coalescing frequencies are the same, the spectra will only depend on the redshift $z$. However, from (\ref{redshift}) one notices that there will be just one value of $\nu_{\mbox{\scriptsize{obs}}}$ for each value of $z$, such that (\ref{bg}) must be handled in a particular way. First, let us rewrite (\ref{bg}) in the form
\begin{equation}
\label{coalNSNS1}
h^{2}_{\mbox{\scriptsize{BG}}}= \frac{1}{\nu_{\mbox{\scriptsize{obs}}}}\int^{z_{p}+\delta z}_{z_{p}-\delta z}h^{2}_{\mbox{\scriptsize{source}}}dR
\end{equation}
where $\delta z<<1$, and $z_{p}$ is the value of redshift corresponding to each observed frequency by means of (\ref{redshift}).

Now, in order to calculate (\ref{coalNSNS1}), $dR$ can be written in the following form
\begin{equation}
\label{coalNSNS2}
dR=\frac{dR}{dV}V(z_{p})\delta(z-z_{p})dz
\end{equation}
where $V(z_{p})$ is the value of the comoving volume at $z_{p}$ and $\delta(z-z_{p})$ is the Dirac's delta function. Therefore, substituting (\ref{coalNSNS2}) in (\ref{coalNSNS1}) and integrating, we have

\begin{equation}
\label{coalNSNS3}
h^{2}_{\mbox{\scriptsize{BG}}}=\frac{1}{\nu\mbox{\scriptsize{obs}}}h^{2}_{\mbox{\scriptsize{source}}}\left.\frac{dR}{dV}\right|_{z_{p},\nu_{obs}}V(z_{p}).
\end{equation}

\subsection{BHNS systems}

The study of the coalescence of BHNS systems is more complicated than the case of NSNS systems, because the frequency of coalescence depends on the mass of the black hole. It is usually assumed that the coalescence occurs when the neutron star reaches the innermost stable circular orbit (ISCO) of the black hole. So, using Kepler's third law and recalling that the emitted frequency is twice the orbital frequency, we have
\begin{equation}
\label{kepler}
\nu=\frac{1}{\pi}\sqrt{\frac{G(m_{\mbox{\scriptsize{ns}}}+m_{\mbox{\scriptsize{bh}}})}{r^{3}_{\mbox{\scriptsize{isco}}}}}
\end{equation}
where $r_{\mbox{\scriptsize{isco}}}$ is the radius of the ISCO of the black hole, which is related to the Schwarzschild radius, namely
\begin{equation}
\label{schw}
r_{\mbox{\scriptsize{isco}}}=3r_{\mbox{\scriptsize{Schw}}}=\frac{6Gm_{\mbox{\scriptsize{bh}}}}{c^{2}}.
\end{equation}
Substituting (\ref{schw}) in (\ref{kepler}) we have
\begin{equation}
\label{massBH}
\nu=\frac{c^{3}}{\pi G\sqrt{216}}\sqrt{\frac{m_{\mbox{\scriptsize{ns}}}+m_{\mbox{\scriptsize{bh}}}}{m^{3}_{\mbox{\scriptsize{bh}}}}}.
\end{equation}
Since we are considering that black holes have masses in the interval  $5\mbox{M}_{\odot}$ $-$ $20\mbox{M}_{\odot}$, the emitted frequencies will have minimum and maximum values given by $\nu^{\mbox{\scriptsize{min}}}=114$Hz and $\nu^{\mbox{\scriptsize{max}}}=997$Hz, respectively. Moreover, considering (\ref{redshift}) and (\ref{massBH}), one notes that for each value of $\nu_{\mbox{\scriptsize{obs}}}$, one has continuous ranges of values for $z$ and $m_{\mbox{\scriptsize{bh}}}$.

However, in the case of BHNS systems, (\ref{bg}) must be integrated over $z$ but not over $m_{\mbox{\scriptsize{bh}}}$, since these variables are not independent. Moreover, for a given value of $\nu_{\mbox{\scriptsize{obs}}}$ one needs to determine the maximum and minimum limits of integration, which are given by
\begin{eqnarray}
z_{\mbox{\scriptsize{min}}}=\frac{\nu^{\mbox{\scriptsize{min}}}}{\nu_{\mbox{\scriptsize{obs}}}}-1\\
z_{\mbox{\scriptsize{max}}}=\frac{\nu^{\mbox{\scriptsize{max}}}}{\nu_{\mbox{\scriptsize{obs}}}}-1.
\end{eqnarray}
When $z_{\mbox{\scriptsize{min}}}<0$ or $z_{\mbox{\scriptsize{max}}}>20$, we set $z_{\mbox{\scriptsize{min}}}=0$ and $z_{\mbox{\scriptsize{max}}}=20$, respectively. Once the interval of integration is set, the variables $m_{\mbox{\scriptsize{bh}}}$ and $\nu$ in the integral can be written as functions of $z$ by solving (\ref{massBH}) and (\ref{redshift}).

In addition, note that in this case one needs to consider in the calculation of (\ref{bg}) the distribution function given by (\ref{blackhole}). As a result, we have
\begin{equation}
%\label{coalesc}
h^{2}_{\mbox{\scriptsize{BG}}}=\frac{1}{\nu_{\mbox{\scriptsize{obs}}}}\int_{z_{\mbox{\scriptsize{min}}}}^{z_{\mbox{\scriptsize{max}}}} h_{\mbox{\scriptsize{source}}}^{2}\frac{dR}{dV}\frac{dV}{dz}f_{1}(m_{1})dz.
\end{equation}

\subsection{BHBH systems}

In this case, the frequency of coalescence depends on the values of the two components of the system. Following Marassi et al.\cite{marassi}, $\nu$ is given by
\begin{equation}
\label{bhbh1}
\nu=\frac{c^{3}}{G}\frac{a_{o}\eta^{2}+b_{0}\eta+c_{0}}{\pi M}
\end{equation}
where $\eta=m_{\mbox{\scriptsize{bh}}_{1}}m_{\mbox{\scriptsize{bh}}_{2}}/M^{2}$ is the symmetric mass ratio and $M$ is the total mass. The polynomial coefficients are $a_{0}=2.974\times 10^{-1}$, $b_{0}=4.481\times 10^{-2}$ and $c_{0}=9.556\times 10^{-2}$.
According to (\ref{bhbh1}), for each value of $\nu$, there will be a continuous set of pairs of values for $m_{\mbox{\scriptsize{bh}}_{1}}$ and $m_{\mbox{\scriptsize{bh}}_{2}}$ that satisfies this equation. Moreover, the masses are not independent, in fact they are related to each other by (\ref{bhbh1}). Therefore, we should integrate over one of the masses (or one parameter which describes both masses.)

However, it is not possible to solve (\ref{bhbh1}) analytically, i.e, to write the masses as functions of $\nu$. Therefore, we need to use approximations. First, consider that $m_{\mbox{\scriptsize{bh}}_{1}}$ and $m_{\mbox{\scriptsize{bh}}_{2}}$ (which we refer as to $m_{1}$ and $m_{2}$) are related to each other by
\begin{equation}
\label{bhbh2}
m_{1}=k m_{2}
\end{equation}
where the variable $k$ is greater than or equal to one. Substituting (\ref{bhbh2}) in (\ref{bhbh1}) we can write the masses $m_{1}$ and $m_{2}$ as functions of $\nu$ and $k$, namely
\begin{eqnarray}\nonumber
\label{bhbh3}
m_{1}&=&k\frac{c^{3}}{G}\frac{1}{\pi \nu}\left[\frac{a_{0}k^{2}}{(1+k)^{5}}+\frac{b_{0}k}{(1+k)^{5}}+\frac{c_{0}}{1+k}\right]\\\nonumber
m_{2}&=&\frac{c^{3}}{G}\frac{1}{\pi \nu}\left[\frac{a_{0}k^{2}}{(1+k)^{5}}+\frac{b_{0}k}{(1+k)^{5}}+\frac{c_{0}}{1+k}\right].
\end{eqnarray}

A suitable approximation for (\ref{bhbh1}) is given by
\begin{equation}
\label{bhbh4}
m_{1}+m_{2}\approx \frac{c^{3}}{G}\frac{c_{0}}{\pi \nu},
\end{equation}
since $\eta<1$ for all values of $m_{1}$ and $m_{2}$. Considering that $m_{2}=m_{\mbox{\scriptsize{min}}}=5\mbox{M}_{\odot}$, we use (\ref{bhbh4}) in order to estimate a first value for $m_{1}$; next, we take this pair of values for the masses and calculate a first value for $\eta$; we then correct the value for $m_{1}$ using
\begin{equation}
\label{bhbh5}
m_{1}+m_{2}\approx \frac{c^{3}}{G}\frac{a_{0}\eta^{2}+b_{0}\eta+c_{0}}{\pi \nu}.
\end{equation}
With this new value of $m_{1}$, we repeat the above process: we calculate $\eta$ again and use (\ref{bhbh5}), bearing in mind that this process may be performed an arbitrary number of times in order to yield more accurate values for $m_{1}$. In the cases where we have $m_{1}>m_{\mbox{\scriptsize{max}}}=20\mbox{M}_{\odot}$ at the end of the process, we consider $m_{1}=m_{\mbox{\scriptsize{max}}}$ and perform an analogous process to find out $m_{2}$.

With the pair $(m_{1},m_{2})$ at hand, we calculate the maximum value of $k$:
\begin{equation}
\label{bhbh6}
k_{\mbox{\scriptsize{max}}}=\frac{m_{1}}{m_{2}}.
\end{equation}

Thus, in order to cover all possible values of $m_{1}$ and $m_{2}$, we consider that $k$, in (\ref{bhbh3}), ranges form one to $k_{\mbox{\scriptsize{max}}}$.

Finally, (\ref{bg}) can be written as follow
\begin{eqnarray}\nonumber
%\label{coalesc}
h^{2}_{\mbox{\scriptsize{BG}}}&=&\frac{1}{\nu_{\mbox{\scriptsize{obs}}}}\int_{1}^{k_{\mbox{\scriptsize{max}}}}\int_{z_{\mbox{\scriptsize{min}}}}^{z_{\mbox{\scriptsize{max}}}} h_{\mbox{\scriptsize{source}}}^{2} \\ &&\times\frac{dR}{dV}\frac{dV}{dz}f_{1}(m_{1})f_{2}(m_{2})dzdk.
\end{eqnarray}

\section{Results and discussion}
\label{sec4}

As usual, in the literature, we represent the backgrounds in terms of the strain amplitude $S_{h}$, which is given by

\begin{equation}
\label{strain}
S_{h}=\frac{h^{2}_{\mbox{\scriptsize{BG}}}}{\nu_{\mbox{\scriptsize{obs}}}}
\end{equation}
and also in terms of the energy density parameter $\Omega_{\mbox{\scriptsize{gw}}}$, which reads (see, e.g., Ref. \cite{ferrari2})

\begin{equation}
\label{omega}
\Omega_{\mbox{\scriptsize{gw}}}=\frac{4\pi^{2}}{3H_{0}^{2}}\nu^{2}h_{\mbox{\scriptsize{BG}}}^{2}.
\end{equation}

\begin{figure}
\rotatebox{-90}{\includegraphics[width=0.30\textwidth]{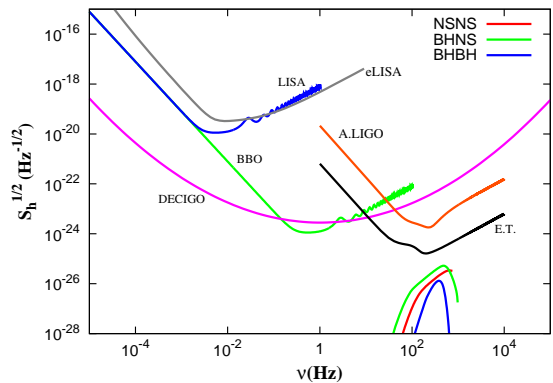}}
\caption{Backgrounds for NSNS, BHNS and BHBH systems in terms of the strain amplitude}
\label{fig1}
\end{figure}

\begin{figure}
\rotatebox{-90}{\includegraphics[width=0.30\textwidth]{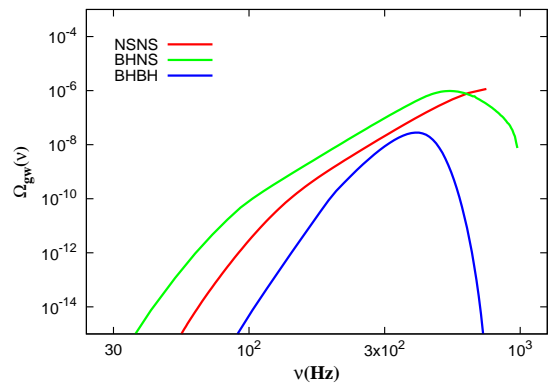}}
\caption{Backgrounds for NSNS, BHNS and BHBH systems in terms of the energy density parameter}
\label{fig2}
\end{figure}

The spectra are shown in Fig.~\ref{fig1} and Fig.~\ref{fig2} for $S_{h}$ and $\Omega_{\mbox{\scriptsize{gw}}}$, respectively. Note that these spectra are mainly compared to the sensitivity curves of ET\cite{satya} and ALIGO\cite{satya}, since the spectra are in their frequency bands. The sensitivity curves of the proposed space-based antennas LISA\footnote{see \url{www.srl.caltech.edu/~shane/sensitivity}}, eLISA\cite{elisa}, BBO\cite{bbo}, DECIGO\cite{decigo} are also shown, but given their frequency bands they cannot detect the spectra studied here.

From Fig. \ref{fig1}, one notices that the background generated by the three families of compact binaries are below the sensitivity curves of the interferometric detectors. Besides, one notices that the background generated by the BHNS systems have higher amplitudes when compared to the ones generated by NSNS and BHBH systems. On the other hand, the spectrum corresponding to BHBH systems has the lowest amplitudes.

Since our calculations depend on some parameters and functions, it is worth investigating how our results are affected by different choices of these quantities.

First, let us consider the masses of the components: in (\ref{source}), if we multiply both masses by a factor of $q$, $h_{\mbox{\scriptsize{source}}}$ will be multiplied by a factor of $\approx q^{1.667}$. Important variations in the amplitudes would occur only if $q<<1$ or $q>>1$.

From (\ref{bg}), (\ref{bfrd}) and (\ref{strain}), one notices that $\sqrt{S_{h}}\propto \sqrt{\dot{\rho}_{\ast}(z_{0})}$ and $\sqrt{S_{h}}\propto \sqrt{\lambda_{\mbox{\scriptsize{nsns}}}}$. Therefore, if ones multiply $\dot{\rho}_{\ast}(z_{0})$ or $\lambda_{\mbox{\scriptsize{nsns}}}$ by, say, a factor of $10$, the amplitudes shown in Fig.~\ref{fig1} will increase by a factor of $\approx 3.2$. Therefore, for realistic scenarios, different choices for $\dot{\rho}_{\ast}(z_{0})$ and $\lambda_{\mbox{\scriptsize{nsns}}}$ would have small effects on the amplitudes of the backgrounds.

Comparing Fig. \ref{fig2} with similar studies found in the literature (see, e.g., Refs. \cite{regim,regim2,regim3,wu,zhu,marassi}), one sees a good agreement concerning the shapes of the spectra, although our results show higher amplitudes. For example, in \cite{regim} one sees that $\sqrt{S_{h}}\sim 10^{-27}\mbox{Hz}^{-1/2}$ at $\nu_{\mbox{\scriptsize{obs}}}\sim 800$Hz for the backgrounds generated by NSNS systems, while for our corresponding spectrum, shown in Fig. \ref{fig1}, we have $\sqrt{S_{h}}\sim 5\times 10^{-26}\mbox{Hz}^{-1/2}$ at the same frequency.

Comparing our results for BHBH systems (see Fig.~\ref{fig2}) with the results found in \cite{zhu}, one can note some similarities: the amplitudes increase until a maximum value in the range $400-500$Hz and then they have a sharp decrease. Concerning the amplitudes, we have a maximum value of $\sim 5\times 10^{-8}$, while in Zhu et al the value is $\sim 2\times 10^{-9}$.

Marassi et al \cite{marassi} also study backgrounds generated by BHBH binaries. These authors discussed different models, and the resulting spectra present maximum amplitudes ranging from  $10^{-10}\leq \Omega_{\mbox{\scriptsize{gw}}} \leq 5\times 10^{-8}$ for a frequency band around  $\sim 500$Hz.

It is worth mentioning that, generally speaking, the spectra is model dependent. Therefore, different assumptions lead to different backgrounds. In Ref.\cite{regim}, for example, the population of binaries is such that the maximum probability of coalescence is around $z=1.4$. Therefore, for $z<1.4$ there is a relatively small proportion of coalescing systems emitting; in our calculations we do not consider such a behavior. This difference in the proportion of systems at lower redshifts could explain our higher amplitudes as compared to the ones of Ref. \cite{regim}.

\subsection{Cross-correlation of pairs of detectors}

Although the spectra (signals) shown in Fig.~\ref{fig1} are below the sensitivity curves of the detectors, it could well be possible detect them by correlating the outputs of two or more detectores. For the correlation of two interferometers, the detectability of a given signal can be quantified by means of the so called signal-to-noise ratio (S$/$N), namely \cite{allen,romano}:

\begin{equation}
\label{SN}
(S/N)^{2}=\left[\left(\frac{9H_{0}^{4}}{50\pi^{4}}\right)T\int \frac{\gamma^{2}(\nu)\Omega_{GW}^{2}(\nu)}{\nu^{6}S_{h}^{1}(\nu)S_{h}^{2}(\nu)}d\nu\right],
\end{equation}
where $S_{h}^{1}$ and $S_{h}^{2}$ are the spectral noise densities, $T$ is the integration time, and $\gamma(\nu)$ is the overlap reduction function, which depends on the relative positions, spatial orientation, and distances of the detectors; and $\Omega_{\mbox{\scriptsize{gw}}}$ is given by (\ref{omega}).

In Table~\ref{tab1} one can see the S$/$N for the three families of compact binaries,
in particular for pairs of ALIGOs and ET.

\begin{table}
\caption{S$/$N for the three families of coalescing compact binaries, considering $T=1$year}
\label{tab1}
\begin{tabular}{lll}
\hline\noalign{\smallskip}
System & ALIGO & ET  \\
\noalign{\smallskip}\hline\noalign{\smallskip}
NSNS & $6.3\times 10^{-3}$ & $5.0\times 10^{2}$ \\
BHNS & $3.4\times 10^{-2}$ & $1.2\times 10^{3}$ \\
BHBH & $1.6\times 10^{-3}$ & $6.4\times 10^{1}$ \\
\noalign{\smallskip}\hline
\end{tabular}
\end{table}

From Table~\ref{tab1}, one notices that ET could in principle detect the backgrounds where the spectrum generated by BHNS systems would have higher probability of detection; for pairs of ALIGOs, the low values of the S$/$N ratio indicate a non detection.

\section{Conclusions}
\label{sec5}

In this paper, we calculate the stochastic background of gravitational waves generated by coalescing compact binaries, using a new method developed in our previous studies \cite{edgard13,edgard14}.

We show that, of the three spectra considered in this paper, the one generated by BHNS systems has the highest amplitudes, while the background by BHBH systems show the lowest amplitudes. Moreover, one notices slight differences in the forms of the spectra, which are due to the different methods used to calculate them by means of (\ref{bg}).

We found that the backgrounds calculated here would not be detected by the interferometric detectors such as LIGO and ET, although thanks to the cross-correlation of signals ET could, in principle, detect such signals.  Particularly, we found that the spectrum generated by BHNS systems have the highest S$/$N ratio, while the one corresponding to BHBH systems presents the lowest S$/$N.

Concerning the dependence of our results on the parameters used in the calculations, we found that the masses of the components of the binaries, as well as $\dot{\rho}_{\ast}(z_{0})$ and $\lambda_{\mbox{\scriptsize{nsns}}}$, do not strongly influence the backgrounds. Besides, a particular choice for the NS EOS does not affect the results either.

We compared the spectra studied here with some interesting results found in the literature. One notices similarities in their shapes, namely: maximum frequencies of $\sim 10^{3}$Hz and maximum amplitudes in the range $400-800$Hz. Roughly speaking, these characteristics are common to the three families of binaries.

On the other hand, our amplitudes given in terms of $\sqrt{S_{h}}$ are in general higher than the ones found in the literature by roughly one order of magnitude. We concluded that such a difference is mainly due to population characteristics assumed.
Therefore, generally speaking, the spectra is model dependents.
% new test above -

\begin{acknowledgements}
EFDE would like to thank Capes for support and JCNA would like to thank FAPESP and CNPq for partial support. Finally, we thank the referee for the careful reading of the paper, the criticisms, and the very useful suggestions which greatly
improved our paper.
\end{acknowledgements}

\appendix

\section{Appendix}
\label{apenA}

In this appendix, we present the main steps of the derivation of $dR/dV$. For further details, we refer the reader to Refs. \cite{edgard13,edgard14}.

We derive $dR/dV$ by means of an analogy with a problem of Statistical Mechanics. In this problem, the aim is to calculate the number of particles that reach a given area $A$ in a time interval $dt$, i.e., the objective is to calculate the flux $F$ of particles. Basically, this flux is calculated by counting the particles inside the volume $dV=Adx$, adjacent to the area $A$, that are moving towards $A$ with velocity $v=dx/dt$, where $v$ obeys a distribution function $\eta(v)$. Hence, the flux is obtained by integrating over all the positive values of $v$.

With some modifications, this method can be used to determine $dR/dV$. First, we substituted the spatial coordinate $x$ by the frequency $\nu$ and the velocity $v$ by the time variation of the frequency, which is defined by $\upsilon_{\nu}=d\nu/dt$. Therefore, the number of systems in the interval $d\nu$ adjacent to a particular frequency $\nu$ is given by

\begin{equation}
    \psi(\nu)=\frac{\varphi(\nu)}{\int \varphi(\nu^{\prime})d\nu^{\prime}}
\label{back2}
\end{equation}
where $\varphi(\nu)$ is the non-normalized distribution of frequencies.

Considering that the distribution $\eta(\upsilon_{\nu})$ gives the number of systems which have $\upsilon_{\nu}$ in the interval $d\upsilon_{\nu}$, the number of systems in $d\nu$ and with values of $\upsilon_{\nu}$ in the interval $d\upsilon_{\nu}$ is given by

\begin{equation}
    d\mu=\left(\frac{\varphi(\nu)d\nu}{\int \varphi(\nu^{\prime})d\nu^{\prime}}\right)\eta(\upsilon_{\nu})d\upsilon_{\nu}.
\label{dmu}
\end{equation}
Now, the next step is to determine $\varphi(\nu)$ and $\eta(\upsilon_{\nu})$. First, the distribution $\varphi(\nu)$ is written in the form

\begin{equation}
\label{inte}
   \varphi(\nu) =\int n_{\mbox{\scriptsize{bin}}}(t_{o})H(\nu)dt_{o}
\end{equation}
where $t_{0}$ is the instant of birth of the systems, $n_{\mbox{\scriptsize{bin}}}$ is the formation rate density of the NSNS, and $H(\nu)$ is given by (\ref{dist2}).

In the derivation of (\ref{inte}), we consider initially $H(\nu)$, from which we have

\begin{equation}
dn=H(\nu)d\nu ,
\end{equation}
which is the fraction of systems originated at the time $t_{0}$ and that have frequencies in the interval $d\nu$. Now, using $n_{\mbox{\scriptsize{bin}}}$, we can write explicitly

\begin{equation}
\frac{dn}{d\nu dVdt_{0}}=n_{\mbox{\scriptsize{bin}}}(t_{0})H(\nu).
\end{equation}
Now, integrating over $dt_{0}$, we get

\begin{equation}
\frac{dn}{dV}=\left[\int n_{\mbox{\scriptsize{bin}}}(t_{0})H(\nu)dt_{0}\right]d\nu ,
\label{app1}
\end{equation}
where the expression in brackets is the number of systems per unit frequency interval and per comoving volume, which is the desired distribution function $\varphi(\nu)$.

On the other hand, $\eta(\upsilon_{\nu})$ will have a peculiar form. First, note that the derivation of (\ref{freqev}) yields

\begin{equation}
\upsilon_{\nu}\equiv \frac{d\nu}{dt}\propto \nu^{\frac{11}{3}}
\end{equation}
after some algebraic manipulations. We conclude that there will be just one value of $\upsilon_{\nu}$ for each value of $\nu$, which allows us to write $\eta(\upsilon_{\nu})$ as a Dirac's delta function, namely

\begin{equation}
\label{xi}
    \eta(\upsilon_{\nu})=N\delta(\upsilon_{\nu}-\upsilon_{\nu,p}),
\end{equation}
where $N$ is the total number of systems and $\upsilon_{\nu,p}$ is the particular value of $\upsilon_{\nu}$ corresponding to each frequency $\nu$.

Notice that the denominator of the term between parenthesis in (\ref{back2}) is the total number of systems. Now, using the function given by (\ref{xi}) and changing the differential $d\nu$ by means of the chain rule, (\ref{dmu}) assumes the form

\begin{equation}
d\mu=\left(\frac{\varphi(\nu)\frac{d\nu}{dt}dt}{N}\right)N\delta(\upsilon_{\nu}-\upsilon_{\nu,p})d\upsilon_{\nu}.
\end{equation}
Integrating over $\upsilon_{\nu}$ and rearranging the terms, we obtain

\begin{equation}
R=\varphi(\nu)\frac{d\nu}{dt}
%d\mu=\left[\varphi(\nu)\frac{d\nu}{dt}\right]dt,
\end{equation}
where $R$ is the number of systems per time interval $dt$. Recalling that the rate $R$ is per comoving volume, one has

\begin{equation}
\label{bg4}
\varphi(\nu)\frac{d\nu}{dt}\equiv \frac{dR}{dV}.
\end{equation}

\section{Appendix}
\label{apenB}

In this paper, the frequency distribution was derived from the semi-major axis distribution given by the following gaussian function\cite{belczynski}

\begin{equation}
f_{\mbox{\scriptsize{G}}}(r)=C\hspace{1pt}\mbox{exp}\left[\frac{-(r-\bar{r})^{2}}{2\sigma^{2}}\right] ,
\label{distr3}
\end{equation}
where $r$ is the semi-major axis and the parameters $\bar{r}$, $C$ and $\sigma$ have the values given in Table \ref{tab2}.

\begin{table}
\caption{Parameters of the distribution functions of the orbital distances}
\label{tab2}
\begin{tabular}{llll}
\hline\noalign{\smallskip}
System & $C$ & $\bar{r}(R_{\odot})$ & $\sigma(R_{\odot})$ \\
\noalign{\smallskip}\hline\noalign{\smallskip}
NSNS & $0.070$ & $0.6$ & $0.2$ \\
BHNS & $0.015$ & $5.5$ & $1.5$ \\
BHBH & $0.070$ & $11$ & $2.5$ \\
\noalign{\smallskip}\hline
\end{tabular}
\end{table}

First, we changed variables via $f(r)dr=g(\Omega)d\Omega$ with the aid of Kepler's third law, where $\Omega$ is the angular orbital frequency. Note that $\Omega$ depends on time (see Ref.\cite{peters}), namely

\begin{equation}
\Omega=\left[\Omega_{0}^{-8/3}-\frac{8}{3}K(t-t_{0})\right]^{-3/8},
\label{freqev}
\end{equation}
where $K=96m_{1}m_{2}5c^{5}G^{5/3}(m_{1}+m_{2})^{-1/3}$, $m_{1}$ and $m_{2}$ are the masses of the components of the system and $\Omega_{0}$ is the initial frequency. So, carrying out a change of variables via $g(\Omega_{0})d\Omega_{0}=H(\Omega)d\Omega$, where $\Omega_{0}$ was associated with the variable $\Omega$ in $g(\Omega)$, one has

\begin{equation}
H(\Omega)=\frac{2C}{3}\left[\frac{GM}{4\pi}\right]^{1/3}\Omega^{-11/3}\Omega_{0}^{2}\mbox{exp}\left[\frac{-(r-\bar{r})^{2}}{2\sigma^{2}}\right].
%\label{dist2}
\end{equation}
Finally, it would be necessary to perform a further coordinate transformation in order to write $H(\Omega)$ as a function of the emitted frequency $\nu$. Such a transformation, calculated by means of $\Omega=\pi\nu$, is trivial. Besides, $r$ is written as a function of $\Omega$ by means of Kepler's third law.

\end{document}